\documentclass{amsart}       

\usepackage{graphicx}
\usepackage{makecell}
\usepackage{mathptmx}      
\usepackage{latexsym}
\usepackage{color}
\usepackage{mathptmx}
\usepackage{url}
\usepackage{bm}
\usepackage{cite}
\usepackage{amsmath}
\usepackage{amssymb}
\usepackage{bbm}
\usepackage{graphicx}
\usepackage{subfigure}
\usepackage{tikz}
\usepackage{array}
\usepackage{multirow, makecell}
\usetikzlibrary{arrows,positioning,shapes.geometric}
\usetikzlibrary{matrix,chains,decorations.pathreplacing}
\usepackage{xcolor}
\usepackage{microtype}
\usepackage{booktabs}
\usepackage{url,diagbox}
\usepackage{comment}
\newcommand{\ud}{\,\mathrm{d}}

\begin{document}

\title[Cadence estimation by dsSST]{Application of de-shape synchrosqueezing to estimate gait cadence from a single-sensor accelerometer placed in different body locations}

\author{Hau-Tieng~Wu}
\address{Hau-Tieng Wu\\
Departments of Mathematics and Department of Statistical Science\\
Duke University, Durham, NC, United States of America}
\email{hauwu@math.duke.edu}

\author{Jaroslaw Harezlak}
\address{Jaroslaw Harezlak\\
Department of Epidemiology and Biostatistics, Indiana University School of Public Health, Bloomington, IN, United States of America
}
\email{harezlak@iu.edu}

\begin{abstract}

{\bf Objective:} 
Commercial and research-grade wearable devices have become increasingly
popular over the past decade. Information extracted from devices using accelerometers is frequently summarized as ``number of steps" (commercial devices) or ``activity counts" (research-grade devices). Raw accelerometry data that can be easily extracted from accelerometers used in research, for instance ActiGraph GT3X+, are frequently discarded. 
{\bf Approach:} 
Our primary goal is proposing an innovative use of the {\em de-shape synchrosqueezing transform} to analyze the raw accelerometry data recorded from a single sensor installed in different body locations, particularly the wrist, to extract {\em gait cadence} when a subject is walking. %
The proposed methodology is tested on data collected in a semi-controlled experiment with 32 participants walking on a one-kilometer predefined course. Walking was executed on a flat surface as well as on the stairs (up and down). 
{\bf Main Results:} 
The cadences of walking on a flat surface, ascending stairs, and descending stairs, determined from the wrist sensor, are 1.98$\pm$0.15 Hz, 1.99$\pm$0.26 Hz, and 2.03$\pm$0.26 Hz respectively. The cadences are 1.98$\pm$0.14 Hz, 1.97$\pm$0.25 Hz, and 2.02$\pm$0.23 Hz, respectively if determined from the hip sensor, 1.98$\pm$0.14 Hz, 1.93$\pm$0.22 Hz and 2.06$\pm$0.24 Hz, respectively if determined from the left ankle sensor, and 1.98$\pm$0.14 Hz, 1.97$\pm$0.22 Hz, and 2.04$\pm$0.24 Hz, respectively if determined from the right ankle sensor. The difference is statistically significant indicating that the cadence is fastest while descending stairs and slowest when ascending stairs. Also, the standard deviation when the sensor is on the wrist is larger. These findings are in line with our expectations. 
{\bf Conclusion:} 
We show that our proposed algorithm can extract the cadence with high accuracy, even when the sensor is placed on the wrist. \\

{\bf Keywords:} actinogram; cadence; de-shape; synchrosqueezing.
\end{abstract}

\maketitle

\section{Introduction}

Wearable physical activity (PA) monitors based on accelerometers have been widely used to characterize the free-living movement of individuals in large scale observational studies, including The National Health and Nutrition Examination Survey (NHANES)\cite{troiano_physical_2008}, UK Biobank \cite{doherty_large_2017}, Women's Health Initiative \cite{evenson_calibrating_2015}, Study of Latinos (SoL) \cite{arredondo_physical_2016}, among many others. Accelerometers are small, non-invasive and body-worn devices that can collect data continuously  in three orthogonal axes for several days at a time and provide high-density measurements of movement ranging from 1 to 100 Hz \cite{karas_accelerometry_2019}. Such ecological data collection is potential for providing PA characteristics that can be considered objective and free of a recall bias compared to the traditional questionnaire-based assessment.

The most commonly considered PA characteristic is unarguably the 
PA volume (or PA level). It has been linked to improved cardiorespiratory fitness \cite{dunn_comparison_1999} and reduced risk of chronic conditions \cite{chakravarthy_obligation_2002}, including cardiovascular disease \cite{thompson_exercise_2003}, stroke \cite{lee_physical_2003}, diabetes \cite{lamonte_physical_2005}, breast \cite{rockhill_prospective_1999} and colon cancers \cite{wolin_physical_2009}, osteoporosis \cite{borer_physical_2005}, depression \cite{strohle_physical_2009}, and declining physical \cite{dipietro_physical_2001} and cognitive functioning \cite{lautenschlager_effect_2008}. While daily PA volumes are important themselves and have been extensively studied, high-fidelity data collected with wearable accelerometers create new and exciting opportunities for healthcarers to obtain more detailed information about PA.
It has been shown in several controlled experiments that accelerometry data can be used to estimate kinematic movement characteristics with particular emphasis on ambulation, as its characteristics, traditionally measured in the controlled clinical environment, are related to many health conditions including Parkinson's disease \cite{bloem_falls_2004} and obesity \cite{ko_characteristic_2010}, and survival in older adults \cite{studenski_gait_2011-1}. 
Among many such characteristics, what we are concerned with in this work is the {\em gait cadence}. The gait cadence is defined as the number of steps taken during a fixed time unit while walking, usually expressed in steps-per-minute or steps-per-second and directly related to gait speed, which is a temporal metric that is well-reflected in high-density accelerometry data \cite{del_din_measuring_2016}.

Accelerometry-based monitoring of gait cadence has successfully migrated from in-the-lab to free-living settings, with the use of devices worn on the lower-back \cite{del_din_analysis_2019}, hip \cite{trojaniello_comparative_2015}, thigh \cite{del_din_free-living_2016}, and ankle \cite{mannini_activity_2013}.
While lab-based assessment of gait cadence has been widely used and well-studied, it is still challenging to implement in real-life settings given the time and space limitations of health facilities. Ideally, we would like to measure mobility with easy-to-access, inexpensive, and multipurpose equipment, like the wrist-based accelerometry measurements.
Clearly, using a watch-like monitor that does not need to be removed for sleep and does not require burdensome body adhesives \cite{del_din_analysis_2019} or bandages \cite{koster_comparison_2016} can increase the participant comfort levels resulting in improved compliance and diminished recruitment hardship \cite{troiano_evolution_2014}. 

Methods dedicated for accelerometry data collected from wrist-located devices are still in active development. For example, relatively simple methods are based on zero-crossings \cite{zijlstra_assessment_2003} and windowed zero-crossings \cite{gonzalez_real-time_2010} and can be used to estimate cadence based on data collected by trunk- and hip-worn devices. Fasel et al. \cite{fasel_wrist_2017} proposed the comb filtering approach to estimate the fundamental frequency of the accelerometry signal observed during walking and related it to the gait cadence, while Karas et al. \cite{karas_adaptive_2019} used dictionary-based pattern recognition to segment walking strides in the accelerometry data to assess the cadence based on the estimated duration of strides. These works, among others, suggest that walking-originated accelerometry data can be modeled within a wide frame of non-stationary time series analysis. In \cite{chinimilli2019two}, the angular data recorded from one inertial motion unit (IMU) attached to the thigh was used to evaluate various activities in the controlled and uncontrolled environment and transitions between activities. Other works that utilize angular motion data recorded by the single IMU worn on the thigh in the controlled environment are \cite{bartlett2017phase,quintero2017real}. In \cite{quintero2017real}, the gait speed is also estimated via the polar radius of the phase portrait. Analogously, Shin and colleagues \cite{shin_adaptive_2011}, proposed to estimate the averaged walking distance from data collected by sensors located on the waist using a linear combination of ``walking frequency'' and ``acceleration variance'' detected by zero crossing. 
While there have been many works, such accelerometry data collected in a free-living environment still impose major methodological and computational challenges \cite{karas_accelerometry_2019}. We recognize the growing need for robust analytical methods for wrist-worn devices, which motivates the focus of this manuscript on wearable accelerometry data, particularly those collected in a semi-controlled and free-living environment.

To illustrate the complexity and non-stationarity of the accelerameter data, we consider an example collected from a wrist-worn accelerometer with a sampling rate of 80 Hz in one adult during short bouts of walking followed by a continuous 400-meter walking session. 
\begin{figure}[htb!]\centering
\includegraphics[trim=0 160 0 50, clip,width=1\textwidth]{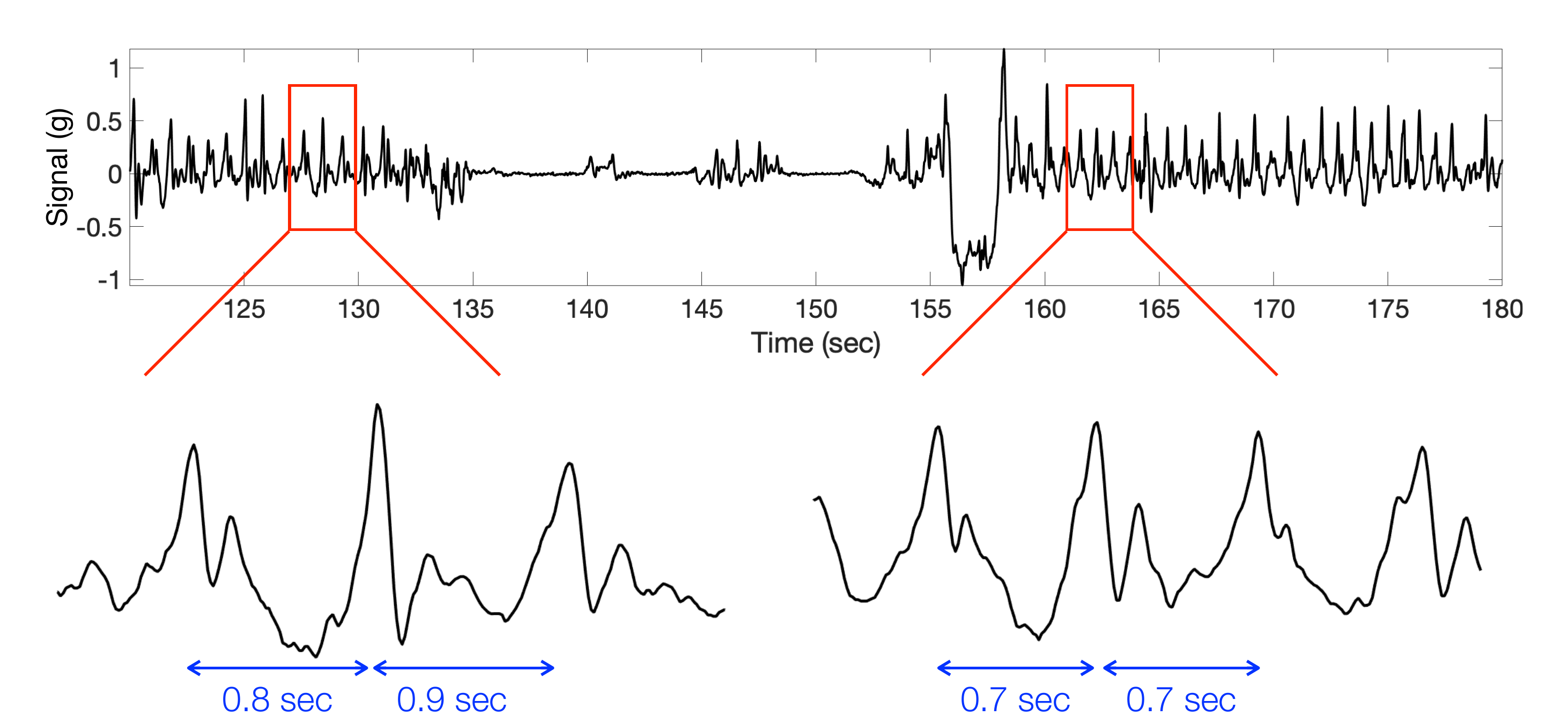}
\caption{A typical accelerometer signal recorded from one subject that performs different physical activities lasting for one minute. The trend has been removed. The subject is walking during the first part and third parts of the signal and is stationary in the middle. The 3-second-long segments were zoomed in and plotted in the bottom panels.  \label{fig1}}
\end{figure}
The top panel of Figure \ref{fig1}  shows an example of the vector magnitude of accelerometry data collected during walking and standing still. The bottom panel represents two instances of zoomed-in strides. We can observe that portions of the signal, corresponding to walking bouts, manifest a quasi-periodic character. Additionally, three rest breaks, between 135 and 153 seconds, can be seen in the form of almost flat lines in the signal, followed by an arm position change between 156 and 158 seconds. Furthermore, both the magnitude and frequency of accelerometry data in walking bouts changes throughout the walk, likely due to the variation in gait cadence. Here we list the most evident non-stationary characteristics of the data that we will further address in our modeling framework:   
\begin{enumerate}
\item[(F1)] The frequency and magnitude of walking accelerometry data undergo smooth variations. 

\item[(F2)] Walking occurs in short-time bouts between resting and non-walking activities. 

\item[(F3)] The walking patterns in accelerometry data are non-sinusoidal and consist of a fundamental frequency with multiple harmonic components. 

\item[(F4)] Accelerometry data contain a random noise component.

\end{enumerate}

The goal of this work is to harness the above-mentioned non-stationary characteristics and faithfully estimate the time trajectory of gait cadence in accelerometry data using time-frequency representation (TFR). To that end, we propose a novel use of the de-shaped synchrosqueezing transform (dsSST), previously introduced by Wu et al. \cite{lin2016waveshape} and develop an algorithm to track the walking cadence (twice of the time-varying fundamental frequency of the accelerometer data). DsSST has been applied to various challenging biomedical signals, like extracting fetal electrocardiogram from the transabdominal maternal electrocardiogram \cite{su2017extract} and recycling cardiogenic artifacts in impedance pneumography \cite{lu2019recycling}. We further demonstrate how the proposed algorithm addresses challenges F1 to F4 and validate our approach for data collected independently with four accelerometers in 32 healthy adults.     

The rest of this manuscript is organized as follows. In Section \ref{Section: model}, we describe the mathematical model of the accelerometer signals. In Section \ref{Section: alg}, we detail the dsSST and the proposed cadence estimation algorithm. In Section \ref{Section:results}, we describe the considered database and report the results of the performed analysis.  Finally, in Section \ref{section:discussion}, we provide the conclusions and the discussion.

\section{Mathematical model}\label{Section: model}

Motivated by characteristics (F1)\textendash (F4), 
we propose the following {\em phenomenological walking model} to model the walking activity in the accelerometer signal. 
We denote the data collected from the triaxial accelerometer signals as $f^{[\texttt{loc}]}_x(t)$, $f^{[\texttt{loc}]}_y(t)$, and $f^{[\texttt{loc}]}_z(t)$, where the placement (location) of the sensor is indicated by \texttt{loc}, and $x,y$ and $z$ indicates the three axes. In this work, we consider the PA signal determined by the signal magnitude, which is defined as
\[
Y^{[\texttt{loc}]}(t)=\sqrt{[f^{[\texttt{loc}]}_x(t)]^2+[f^{[\texttt{loc}]}_y(t)]^2+[f^{[\texttt{loc}]}_z(t)]^2}\,.
\]
Below, we ignore the superscript [\texttt{loc}] when we model the signal.

First, we set up notation to describe the ``walking bout''. Assume there are $L\in \mathbb{N}$ non-overlapping nontrivial intervals $I_1,\ldots I_L\subset \mathbb{R}$ that the subject is walking.
The PA signal determined from the accelerometer is modeled by
\begin{equation}\label{Model:equation}
Y(t) = g(t)+\Phi(t),
\end{equation}
where $g(t)$ is the deterministic signal describing the walking activity and $\Phi$ is the inevitable random noise. We make the following assumptions about $g$ and $\Phi$. Fix $\epsilon>0$ a small constant.
\begin{equation}\label{Model:equation2}
g(t) = \sum_{l=1}^L a_{l}(t)s_{l}(\phi_{l}(t))\chi_{I_l}\,,
\end{equation}
where $\chi_{I_l}$ is the indicator function (that is, $\chi_{I_l}(t)=0$ when $t\notin I_l$ and $\chi_{I_l}=1$ when $t\in I_l$) and for each $l=1,\ldots,L$,
\begin{itemize}
\item[(C1)] $\phi_{l}(t)$ is a $C^2$ function that is strictly monotonically increasing denoting the {\em phase function} of the $l$-th walking interval;

\item[(C2)] $\phi_{l}'(t)>0$ is the instantaneous frequency (IF), or {\em momentary stride speed}, of the $l$-th walking interval so that and $|\phi''_l(t)|\leq \epsilon \phi'_l(t)$ for all $t\in I_l$;

\item[(C3)] $a_{l}(t)>0$ is a $C^1$ function denoting the {\em amplitude modulation (AM)} of the $l$-th walking interval so that $|a_l'(t)|\leq \epsilon \phi'_l(t)$ for all $t\in I_l$;

\item[(C4)] $s_{l}$ is a continuous $1$-period function so that $|\hat{s}_l(1)|>0$ and $\|s_l\|_2=1$, which may not be sinusoidal; 
\end{itemize}
and 
\begin{itemize}
\item[(C5)] $\Phi(\cdot)$ is a random noise that is stationary in the wide sense with finite variance and short range dependence.
\end{itemize}

The cadence is defined as the double of the momentary speed of stride; that is, $2\phi_l'(t)$ over $I_l$. Note that compared with the traditional consideration of cadence that is measured
steps-per-minute or steps-per-second from the high-density accelerometry data \cite{del_din_measuring_2016}, in this work we consider the cadence in the {\em instantaneous} sense. Plainly, the cadence is defined on the sub-second level, which captures the fact that the walking speed changes over time, even within one stride.  
Physically, $a_l$ means the ``strength'' of the walking activity and
$s_l$  is understood as the {\em wave-shape function} describing the walking pattern of the $l$-th walking interval. Both strength and pattern might change from one step to another. Obviously, since the walking activity is a synergy of different activities in different body locations, $a_l$ and $s_l$ depend on the sensor location. 
The phase $\phi_l$ deserves some more discussion. While physically $\phi_{l}'(t)$ should be the same across different sensor locations, the phase $\phi_l$ might be different due to the global phase shift \cite{alian2022reconsider}; that is, the phase of the $l$-th walking bout recorded from two different locations might be different by a global constant, while their derivatives are the same.
On the other hand, even over the same sensor location, $s_{l}$, $\phi_l(t)$, $\phi_{l}'(t)$ and $a_{l}(t)$ might be different for different walking bout. For example, the walking pattern on the road during $I_i$ might be different from the stair climbing pattern during $I_j$, even if the signal is recorded from the hip sensor. 

We shall mention that the assumptions $|a_l'(t)|\leq \epsilon \phi'_l(t)$ and $|\phi''_l(t)|\leq \epsilon \phi'_l(t)$ for all $t\in I_l$ for $\epsilon>0$ a small constant in C2 and C3 are for the sake of identifiability of the model and carrying out a theoretical analysis. Since theoretical development is not the focus of this paper, we refer readers with interest to \cite{DaLuWu2011,Chen_Cheng_Wu:2014} for technical details.

Finally, we put an assumption on the lengths of walking intervals. To our knowledge, there does not exist a general consensus about the definition of ``walking'' \cite{urbanek_prediction_2018}. Precisely, should we consider a subject is ``walking'' if he/she only moves one or two steps? In other words, how many continuous steps should exist before we consider the subject is walking over that interval? We do not plan to provide a solution in this work. Instead, we provide a definition based on the property of time-frequency (TF) analysis, like short-time Fourier transform (STFT). Empirically, to stably estimate the instantaneous frequency from an oscillatory signal by STFT, the window function should span long enough to cover $8$-$10$ oscillatory cycles. Thus, only if a subject moves continuously for more than $10$ steps shall we call such process a walking process. Mathematically, this definition is imposed by the following assumption: 
\begin{itemize}
\item[(C6)]
For each $i=1,\ldots,L$, we have $|I_i|>\delta_i$, where $\delta_i>0$ is assumed to be long enough to cover at least $10$ walking cycles. 
\end{itemize}
Note that by definition of $\phi_i'(t)$, it means that $\delta_i> \phi_i^{-1}(10\times 2\pi)-\phi_i^{-1}(0)$, since when a subject finishes one walking cycle the phase grows by $2\pi$. Depending on the cadence, $\delta_i$ clearly vary; the faster the cadence over $I_i$ is, the shorter the $\delta_i$ will be.  

Clearly, $Y(\cdot)$ is a random process. We call the random process satisfying (C1)\textendash(C6) the {\em phenomenological walking model}. In practice, we only obtain one realization of $Y$ (or recorded accelerometer signal) with the main goal of  estimating the cadence over each walking interval from the recorded signal.
It is worth noting that the phenomenological walking model is a generalization of the {\em adaptive non-harmonic model} \cite{Wu:2013,lin2016waveshape} that captures the walking bout effect.  The main signal processing mission in this paper is estimating the cadence $2\phi_l'(t)$ from one noisy realization of $Y(t)$ over $I_l$. Moreover, if $I_l$ is unknown, we need to estimate it.
Last but not the least, we shall mention that this definition is consistent with the definition of {\em sustained harmonic walking} considered in \cite{urbanek_prediction_2018}, where the sustained harmonic walking is defined as {\em walking for at least 10s} (aligned with C6) with {\em low variability of step frequency} (aligned with C2). 

Let us take a deeper look at the wave-shape function (the walking pattern). Assume $L=1$ and $I_1=\mathbb{R}$ to simplify the discussion. Under this assumption, we have
\begin{equation}
f(t)=a_{1}(t)s_{1}(\phi_{1}(t))=a_1(t)\alpha_0+\sum_{j=1}^\infty (\alpha_ja_1(t))\cos(2\pi j\phi_1(t)+\beta_j)\,,\label{Expansion f Fourier series}
\end{equation}
where $\alpha_0=\int_0^1s_1(t)dt$, and $\{\alpha_j\}\subset\mathbb{R}_{\geq0}$ and $\{\beta_j\}\subset[0,2\pi)$ are from Fourier coefficients of $s_1$. 
Usually, we call $(\alpha_1a_1(t))\cos(2\pi \phi_1(t)+\beta_1)$ the {\em fundamental component} of $f(t)$, and for $j\geq 1$, we call $(\alpha_ja_1(t))\cos(2\pi j\phi_1(t)+\beta_j)$ the {\em $j$-th harmonic} of $f(t)$. In the spectrogram (and the associated TFRs determined by the synchrosqueezing transform (SST)) shown in Figure \ref{fig2}, we see that the fundamental component and the harmonics are represented as horizontal curves (indicated by arrows). Visually we can see that both signals oscillate about once per second, while their oscillatory patterns are different. The difference of oscillatory pattern is reflected in the intensities of the harmonics. For the signal in Figure \ref{fig2}(a), we see that $\alpha_1$ is smaller than $\alpha_j$ for any $j>1$ in Figure \ref{fig2}(b). However, for the signal in Figure \ref{fig2}(d), $\alpha_1$ is stronger than other $\alpha_j$ for all $j>1$, which we can see via the intensities of the horizontal curves in Figure \ref{fig2}(e). It is clear that different walking patterns could lead to vary different spectrograms.

\section{Proposed algorithm}\label{Section: alg}

\begin{figure}[htb!]\centering
\includegraphics[trim=40 10 20 40, clip,width=1\textwidth]{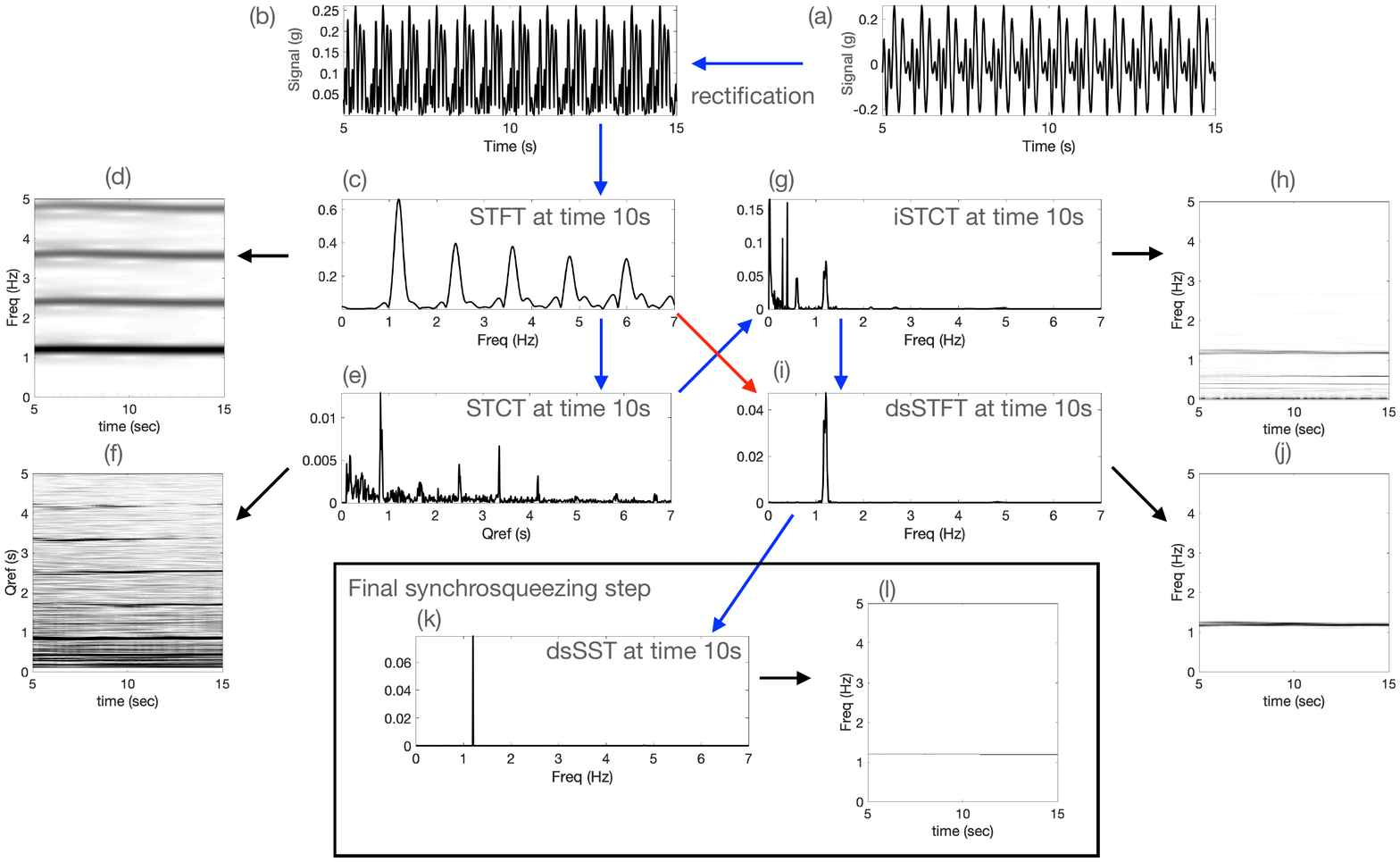}
\caption{The overall flowchart of the de-shape algorithm. (a) is the input signal. (b) is the rectified signal. (c) is the slice of the spectrogram of (b) at the 10th sec, and (d) is the spectrogram. (e) is the slice of the short-time cepstral transform (STCT) of (b) at the 10th sec, and (f) is the STCT. (g) is the inverse STCT (iSTCT) at the 10th sec, which when combined with the spectrogram leads to the de-shape STFT (dsSTFT) at the 10th sec. The associated iSTCT and dsSTFT are shown in (h) and (j) respectively. If the synchrosqueezing transform (SST) is applied, the result at the 10-th sec is shown in (k), and the de-shape SST (dsSST) is shown in (l). \label{figAlgo}}
\end{figure}

In this section, we describe our proposed cadence estimation algorithm.  The overall flowchart of the proposed algorithm is shown in Figure \ref{figAlgo}. 
We shall briefly describe the motivation behind the dsSST development before introducing the algorithm. When the signal oscillates non-sinusoidally, as discussed above with Figure \ref{fig2}, there is no guarantee that the fundamental component is stronger than its multiples, even if the rectification technique \cite{steinerberger2022fundamental} is applied. As a result, it is often difficult to directly apply the ridge detection method \cite{delprat1992asymptotic,carmona1999multiridge} to the TFRs. We thus want a reliable method to filter out its harmonics.
The dsSST contains as key ingredients two nonlinear operators for this purpose -- short-time cepstrum and SST \cite{lin2016waveshape}. In short, cepstrum can be viewed as the Fourier transform of the ``manipulated Fourier transform'', which reflects the period of the oscillation. Given the reciprocal relationship between the frequency and period, the fundamental frequency could be preserved by the masking technique. 
SST is a nonlinear-type TF analysis tool. Its main purpose is sharpening the TFR determined by STFT or CWT for the sake of improving the readability of oscillatory signals by nonlinearly twisting the phase information \cite{DaLuWu2011,Chen_Cheng_Wu:2014}.

\subsection{Cadence estimation algorithm}

\subsubsection{Step 0: Input signal and preprocessing}
Take the PA signal that is uniformly sampled over a discrete set of time points with the sampling interval $\Delta_t>0$ and the sampling rate $f_s = \Delta_t^{-1}$. Suppose the recording starts at time $t = 0$.  Write the uniformly sampled signal as a column vector $\mathbf{f}_0  \in \mathbb{R}^N$, where $N$ is the number of samples and the $\ell$-th entry of $\mathbf{f}_0$ is the signal sampled at time $\ell\Delta_t$, where $\ell=1,2,\ldots,N$.
Then, detrend the signal by the standard median filter with the order $10f_s$, rectify the detrended signal (that is, take the magnitude of each sample), and denote the resulting signal as $\mathbf{f}$. 
In the real life scenario, it is usually unknown when a subject is walking. To apply the proposed cadence estimation algorithm, we need to estimate $I_l$ in the phenomenological walking model described in Section \ref{Section: model}. We suggest to apply the comb filter based algorithm \cite{urbanek_prediction_2018} to achieve this goal, and continue with the following algorithm.

\subsubsection{Step 1: Short time Fourier transform (STFT)}

Choose a discrete window function $\mathbf{h} \in \mathbb{R}^{2K+1}$, a discretization of a chosen window $h$, which satisfies $\mathbf{h}(K+1) = 1$. Define $2K + 1$ to be the window length. Write $\mathbf{h}' \in \mathbb{R}^{2K+1}$ for the discretization of the derivative of the window function. For example, a discrete Gaussian window (and its derivative) with standard deviation $\sigma>0$ sampled over the interval $[-0.5, 0.5]$ at a sampling interval of $\frac{1}{2K}$ is defined as,
\begin{gather}
\mathbf{h}(k) = e^{\frac{-\left(\frac{k-1}{2K} - 0.5 \right)^2}{2\sigma^2}}\,,\quad
\mathbf{h}'(k) = -\left(\frac{k-1}{2K} - 0.5 \right)\frac{\mathbf{h}(k)}{\sigma^2},
\end{gather}
where $k = 1, \ldots, 2K+1$.
Introduce the parameter $M$ so that $2M$ is the chosen number of points on the frequency axis of our time-frequency representation. 
Evaluate the STFT of $\mathbf{f}$, a matrix $\mathbf{V}_\mathbf{f} \in \mathbb{C}^{N \times (M+1)}$ with entries %
\begin{equation}\label{definition STFT discretization}
\mathbf{V}_\mathbf{f}(n, m) = \sum_{k=1}^{2K+1} \mathbf{f}(n + k - K - 1) \mathbf{h}(k) e^{\frac{-i2\pi (k-1)(m - 1)}{2M}},
\end{equation}
where $\mathbf{f}(l) := 0$ when $l < 1$ or $l > N$, $n=1,\ldots,N$ is the time index and $m=1,\ldots,M+1$ is the frequency index. The spectrogram is thus an $N\times (M+1)$ matrix where the $(n,m)$-th entry is $|\mathbf{V}_\mathbf{f}(n,m)|^2$. See Figure \ref{fig2} for two illustrative examples of a spectrogram. 

\subsubsection{Step 2: Short-time cepstral transform and inverse short-time cepstral transform}

The short-time cepstral transform (STCT) of $\mathbf{f}$ is represented by a matrix $\mathbf{C}_\mathbf{f} \in \mathbb{C}^{N\times 2M}$ with entries
\begin{equation}\label{Equation STCT equation}
\mathbf{C}_\mathbf{f}(n, m') = \sum_{m=1}^{2M} \vert \mathbf{V}_\mathbf{f}(n, m) \vert^\gamma e^{\frac{-i 2\pi (m-1)(m'-1)}{2M}},
\end{equation}
where $\gamma>0$ is the chosen power parameter and $m'=1,\ldots,2M$ is the quefrency index.  We crop $\mathbf{C}_\mathbf{f}$ and consider only the first $M + 1$ columns associated with the positive quefrency axis.
The inverse STCT (iSTCT) of $\mathbf{f}$ is represented by a matrix $\mathbf{U}_\mathbf{f} \in \mathbb{R}^{N \times (M+1)}$. For each time index $n$, consider the function $g_n \colon [0, \infty] \rightarrow \mathbf{R}$ whose known values are
\begin{gather}
g_n\left( \frac{1}{m-1} \right) = \mathbf{C}_\mathbf{f}(n, m) \quad m = 1, ..., M+1.
\end{gather}
The entries of the inverted STCT are calculated by interpolation:
\begin{gather}
\mathbf{U}_\mathbf{f}(n, m) = g_n\left( \frac{m-1}{2M} \right).
\end{gather}
See Figure \ref{figAlgo} for an example of STCT and iSTCT. 

\subsubsection{Step 3: De-shape STFT}

The dsSTFT of $\mathbf{f}$, a matrix $\mathbf{W}_\mathbf{f} \in \mathbb{C}^{N \times (M+1)}$, is given by the pointwise product
\begin{equation}
\mathbf{W}_\mathbf{f}(n, m) = \mathbf{V}_\mathbf{f}(n,m) \mathbf{U}_\mathbf{f}(n, m)\,,
\end{equation}
where $n=1,\ldots,N$ and $m=1,\ldots, M+1$. See Figure \ref{figAlgo} for an example of dsSTFT.

\subsubsection{Step 4: Sharpen STFT and dsSTFT.}

To sharpen the TFR determined by STFT or dsSTFT, we calculate the {\em frequency reassignment rule}. First, calculate 
\begin{equation}
\mathbf{V}'_\mathbf{f}(n, m) := \sum_{k=1}^{2K+1} \mathbf{f}(n + k - K - 1) \mathbf{h}'(k) e^{\frac{-i2\pi (k-1)(m - 1)}{2M}}.
\end{equation}
Choose a threshold $\upsilon > 0$ and calculate the reassignment operator
\begin{gather}\label{definition Omega}
\bm{\Omega}_{\mathbf{f}}^{\upsilon}(n, m) = \left\{
\begin{array}{ll}
-\Im\frac{\mathbf{V}_\mathbf{f}'(n,m)}{\mathbf{V}_\mathbf{f}(n,m)} \frac{N}{2\pi(2K + 1)}&\mbox{ when }|\mathbf{V}_\mathbf{f}(n,m)|> \upsilon\\
-\infty&\mbox{ when }|\mathbf{V}_\mathbf{f}(n,m)|\leq \upsilon.
\end{array}
\right.
\end{gather}
The traditional SST of $\mathbf{f}$, denoted as $S\mathbf{V}_\mathbf{f}^\upsilon\in \mathbb{C}^{N \times (M+1)}$, is implemented by 
\begin{gather}
S\mathbf{V}_\mathbf{f}^\upsilon(n, m) = \sum_{l;\, m= l- \bm{\Omega}_{\mathbf{f}}^{\upsilon}(n, m)} \mathbf{V}_\mathbf{f}(n, l) .
\end{gather}
The dsSST of $\mathbf{f}$, a matrix $S\mathbf{W}_\mathbf{f} ^\upsilon\in \mathbb{C}^{N \times (M+1)}$, is given by the formula
\begin{gather}
S\mathbf{W}_\mathbf{f}^\upsilon(n, m) = \sum_{l;\, m= l-\bm{\Omega}_{\mathbf{f}}^{\upsilon}(n, m)} \mathbf{W}_\mathbf{f}(n, l).
\end{gather}
See Figure \ref{fig2} for an illustration of SST. The weak line in the spectrogram associated with the fundamental component in Figure \ref{fig2}(b) is clearly enhanced in Figure \ref{fig2}(c). In the illustration of dsSST in Figure \ref{fig3}, we can easily see that the harmonics are gone.

\subsubsection{Step 5: Cadence extraction}

In the last step, we estimate the cadence over an interval $I_i$ that the subject is walking. Assume the subject is walking over the interval spanning from $1\leq N_1$ to $N_1<N_2\leq N$. We first fit a curve to the dsSST via
\begin{align}
c^*=&\,\max_{c\in Z_{M+1}^{N_2-N_1+1}}\left(\sum_{m=1}^{N_2-N_1+1}\log\left[\frac{|S\mathbf{W}_\mathbf{f}^\upsilon(c(m),m)|}{\sum_{i=1}^n\sum_{j=1}^{N_2-N_1+1} |S\mathbf{W}_\mathbf{f}^\upsilon(j,i)|}\right]  -\lambda \sum_{m=2}^{N_2-N_1+1} |c(m)-c(m-1)|^2\right)\,,\label{CurveExtractionFormula}
\end{align}
where $Z_{M+1}=\{1,2,\ldots,M+1\}$ and $\lambda> 0$ is the penalty term controlling the regularity of the curve $c$. 
Here, $c\in Z_{M+1}^{N_2-N_1+1}$ indicates a curve in the TFR $S\mathbf{W}_\mathbf{f}^\upsilon$ restricted on the interval spanning from $N_1$ to $N_2$ and $\sum_{m=2}^{N_2-N_1+1}|c(m)-c(m-1)|^2$ quantifies the regularity of $c$. Based on the robustness property of SST \cite{Chen_Cheng_Wu:2014}, the extracted curve $c^*$ is a robust estimator of the IF of the fundamental component $\phi'_1(t)$.
As a result, $\phi'_1$ at time $(N_1+\ell-1)\Delta_t$ is estimated as
\[
\widehat{\phi}_1'((N_1+\ell-1)\Delta_t):=\frac{f_s}{2M}c^*(\ell)\,,
\]
where $\ell=1,\ldots,N_2-N_1+1$ and $\frac{f_s}{2M}$ is the discretization interval in the frequency axis. The cadence at time $(N_1+\ell-1)\Delta_t$ is then estimated as 
\[
\widehat{\texttt{cadence}}((N_1+\ell-1)\Delta_t):=2\widehat{\phi}_1'((N_1+\ell-1)\Delta_t)\,.
\]
A similar procedure could be applied to estimate the cadence over other walking intervals.

\section{Results}\label{Section:results}

\subsection{Database}\label{Section: measurements}
We consider the Indiana University Walking and Driving Study (IUWDS). Data were collected in 2015 on $n=32$ ($19$ females) physically healthy individuals in a wide age range between 21 and 51 years of age in a partially controlled environment.  The individuals were given instructions for a walking route during the experiment that included walking on a flat surface and climbing stairs up and down, but otherwise the behavior was unrestricted. Data were collected using four tri-axial ActiGraph GT3X+ accelerometers at a frequency of 100 Hz. An example of the collected data is presented in Figures \ref{fig1} and \ref{fig2}.
The study was approved by the Institutional Review Board (IRB) of the Indiana University, and all participants provided written informed consent to participate. 
In this database, there are four locations that the tri-axial accelerometer signals are simultaneously recorded from, including wrist (\texttt{wr}), hip (\texttt{hi}), left ankle (\texttt{la}), and right ankle (\texttt{ra}); that is, we have $Y^{[\texttt{wr}]}(t)$, $Y^{[\texttt{hi}]}(t)$ $Y^{[\texttt{la}]}(t)$ and $Y^{[\texttt{ra}]}(t)$ recorded simultaneously from each subject. We only focus on the data with experts' annotation.

\subsection{Parameter choices}
We choose the Gaussian window when we run STFT and dsSST with the window span of 12 seconds. Canonical frequency bins are chosen in the discretization of STFT; that is, $M=N/2$ is chosen in \eqref{definition STFT discretization} and $\gamma=0.3$ in \eqref{Equation STCT equation} and $\upsilon$ in \eqref{definition Omega} is set to $10^{-9}$. The penalty term in \eqref{CurveExtractionFormula} is chosen to be $\lambda=1$. 
These parameters are chosen based on physiological needs and not optimized by any grid search, and the results are not sensitive to a small perturbation of these parameters (results not shown).

\subsection{Visualization}
In Figure \ref{fig3}, we illustrate a typical example of $Y^{[\texttt{ra}]}(t)$ from one of the study participants, where we show its spectrogram, SST and dsSST. The experts' labels are superimposed as the red curve on the presented signal, with "1`` denoting walking, "2`` denoting descending stairs, "3`` denoting ascending stairs, and "0`` denoting clapping, separating different walking activities.
It can be clearly seen that there is a dominant curve around 1Hz in the spectrogram, SST and dsSST, when the subject is walking. As described in the methodology section, physically, the dominant curve indicates the IF of the fundamental component, which quantifies the ``momentary speed of strides'', and the doubled value of this IF is the cadence. The multiples are inevitable in the spectrogram and SST, while the multiples disappear in the dsSST. 
When the subject ascends or descends stairs, the IF fluctuates around 1 Hz and the TFR is blurred; when the subjects clap their hands, the IF is totally blurred. This suggests that the accelerometry signal and its spectral content during walking are different from those during other activities.
The preceding observation agrees with the intuition that when a subject is walking on a flat surface, the cadence is more stable than when ascending or descending stairs. 
To appreciate the challenge caused by the feature (F4), we show another typical example of $f^{[\texttt{wr}]}(t)$ in Figure \ref{fig3-2}. In this case, we see that the strength of fundamental frequency multiples is greater than the fundamental component on the spectrogram. This challenge is alleviated after applying the de-shape algorithm.

\begin{figure}[htb!]\centering
\includegraphics[trim=80 300 100 20, clip,width=1\textwidth]{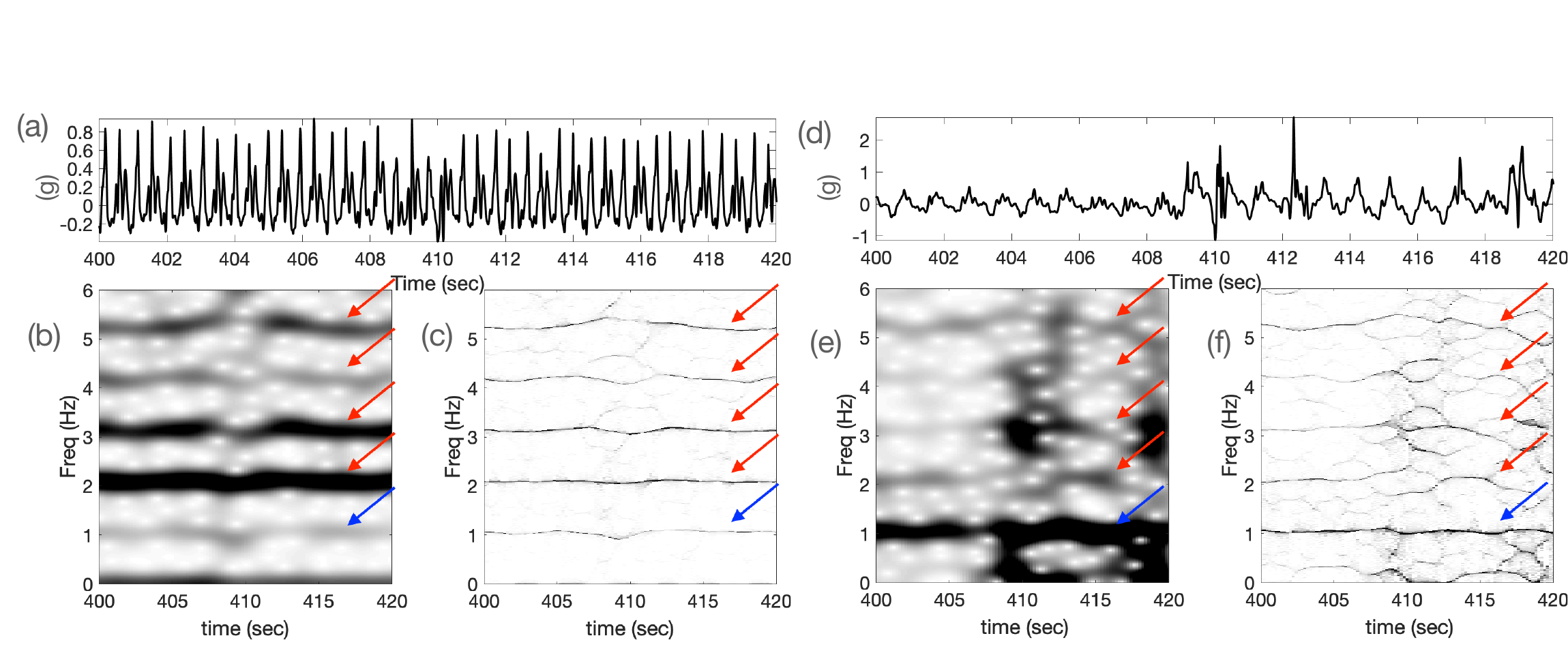}
\caption{The accelerometer signals in panels (a) and (d) were recorded from two subjects. The spectrograms of these signals are shown in panels (b) and (e), respectively and the synchrosqueezed spectrograms of these signals are shown in (c) and (f) respectively.  In panels (b), (c), (e) and (f), the fundamental component is indicated by the blue arrow, and its multiples are indicated by the red arrows. \label{fig2}}
\end{figure}

\begin{figure}[htb!]\centering
\includegraphics[width=1\textwidth]{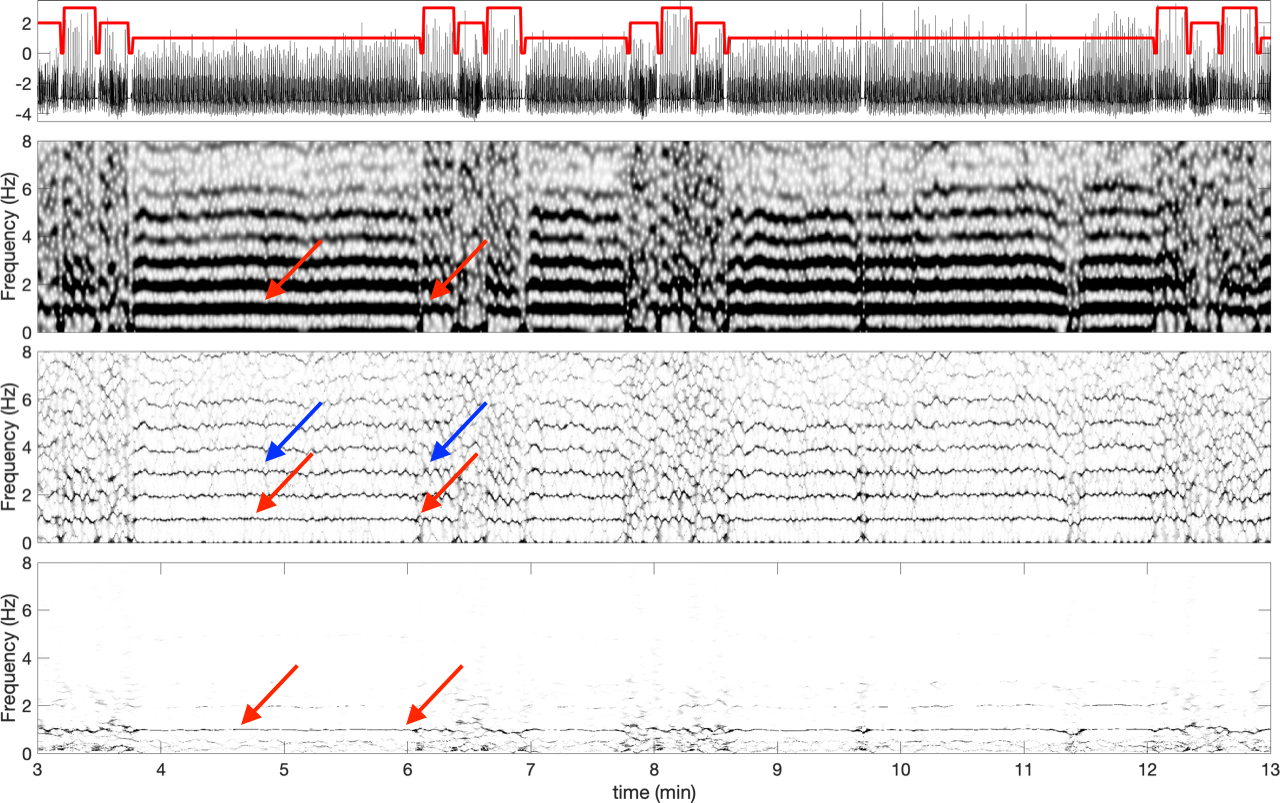}
\caption{From top to bottom: $f^{[\texttt{ra}]}(t)$ superimposed the labels, the spectrogram, the SST, and the dsSST. The red arrow indicates the instantaneous frequency (IF) of the fundamental component (the cadence is the double of the IF), and the blue arrow indicates the IF of the multiple. It is clear that the multiples are suppressed in the dsSST.  To enhance the visualization, the dynamical range of each plot is set to be between $0$ and the $99\%$ quantile of all entries. \label{fig3}}
\end{figure}

\begin{figure}[htb!]\centering
\includegraphics[width=1\textwidth]{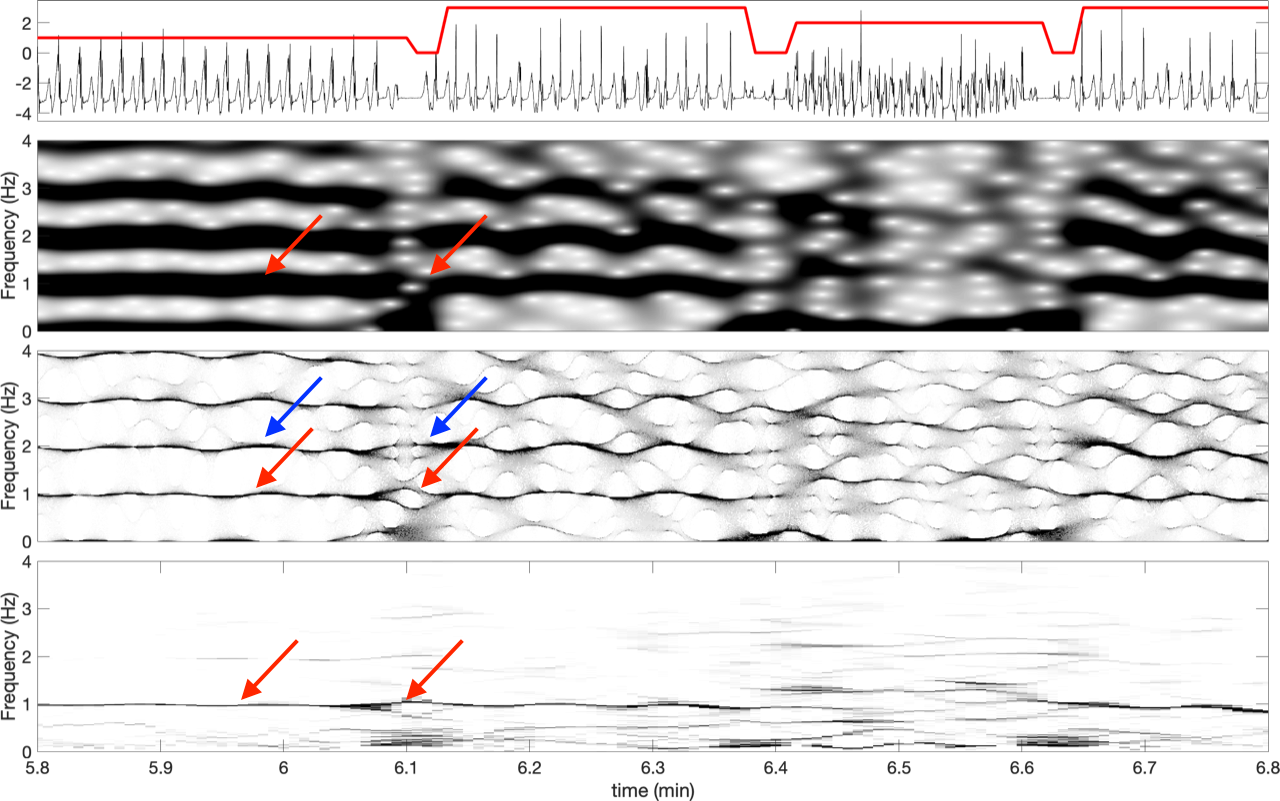}
\caption{A zoom-in of Figure \ref{fig3}. From top to bottom: $f^{[\texttt{ra}]}(t)$ superimposed on the labels, the spectrogram, the SST, and the dsSST. The red arrow indicates the instantaneous frequency (IF) of the fundamental component (the cadence is the double of the IF), and the blue arrow indicates the IF of the multiple. It is clear that the multiples are suppressed in the dsSST.  To enhance the visualization, the dynamical range of each plot is set to be between $0$ and the $99\%$ quantile of all entries. \label{fig4}}
\end{figure}

\begin{figure}[htb!]\centering
\includegraphics[width=1\textwidth]{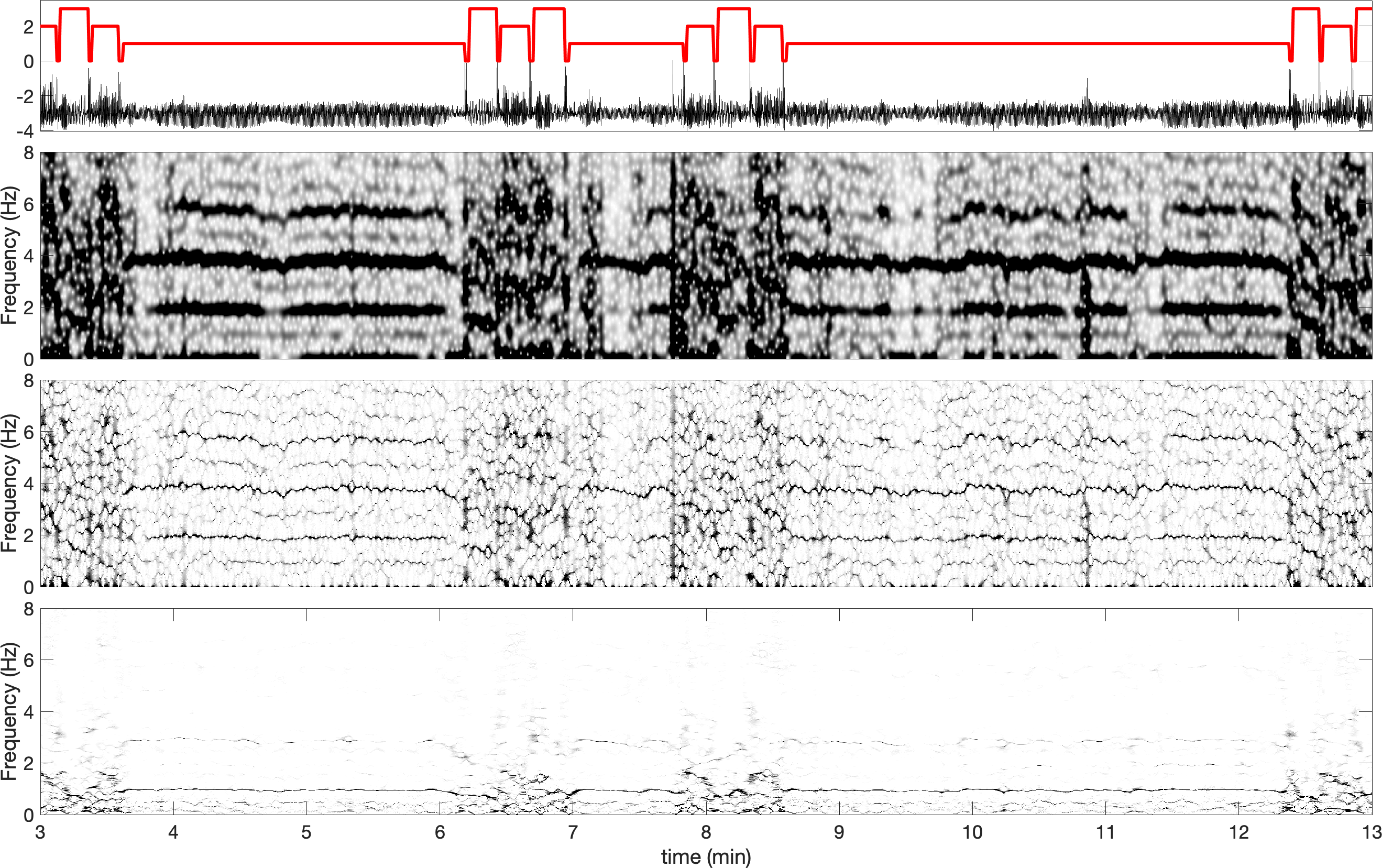}
\caption{From top to bottom: $f^{[\texttt{wr}]}(t)$ superimposed on the labels, the spectrogram, the SST, and the dsSST. The red arrow indicates the instantaneous frequency (IF) of the fundamental component (the cadence is the double of the IF), and the blue arrow indicates the IF of the multiple. It is clear that the multiples are stronger than the fundamental components, and they are suppressed in the dsSST. To enhance the visualization, the dynamical range of each plot is set to be between the $0$ and the $99\%$ percentiles of all entries. \label{fig3-2}}
\end{figure}

\subsection{Statistical analysis of the obtained results}

To analyze data from all study participants, we quantitatively evaluate the cadence over different walking intervals, where the walking intervals are provided by the experts' labels. We set $p=0.005$ \cite{benjamin2018redefine} as the statistical significance level in the following statistical inference and apply the Bonferroni correction if multiple testing is carried out. The results are shown in Table \ref{table0}. 

First, we compare the cadence of different walking activities. For each sensor location, the estimated cadence of ascending stairs is lower than that of walking  on the flat surface, and further lower than descending stairs, where the both differences are with statistical significance under the one-sided Wilcox rank sum test, except comparing the cadence during descending stairs and walking on the flat surface from the hip sensor ($p=0.027$).
Also, when a subject is walking on the flat surface, the standard deviation of the cadence is the smallest compared with ascending and descending stairs with statistical significance under the F test.

Second, we compare different sensor locations. The one-way analysis of variance (ANOVA) shows that during walking on the flat surface, there is no difference ($p=0.377$) when the cadence is estimated from different sensor locations. However, during ascending stairs and descending stairs, there is at least one sensor location that leads to significantly different cadence estimation. This suggests that when a subject is climbing stairs, the cadence estimation agreement among different locations might be lower.
The variability of cadence estimation from the wrist is larger with statistical significance under the F test.

\begin{figure}[htb!]\centering
\includegraphics[width=1\textwidth]{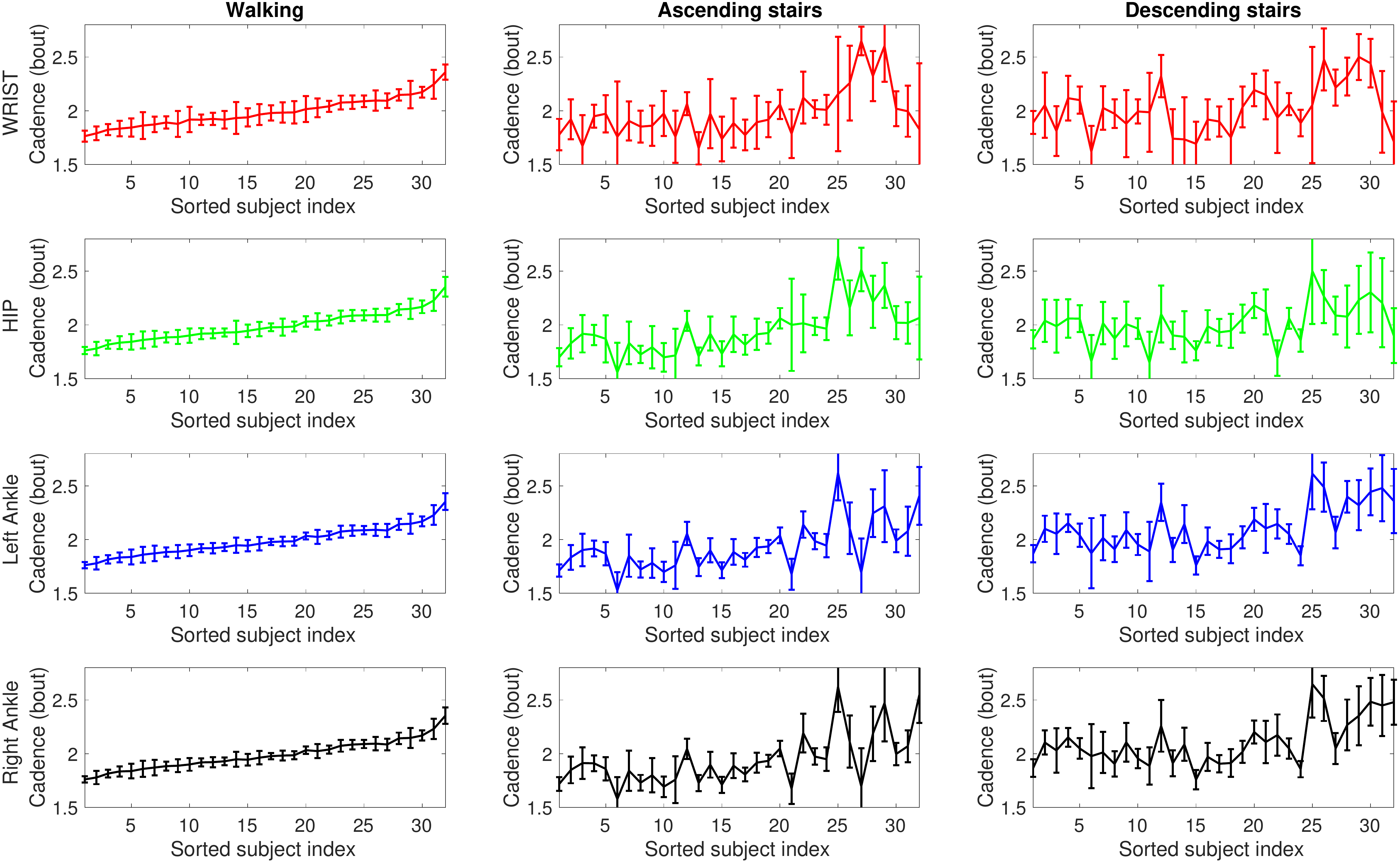}
\caption{A summary plot of means and standard deviations of 32 subjects. The subjects are sorted by the cadence measured from the hip sensor. From left to right: when a subject is walking, ascending stairs and descending stairs. From top to bottom: when the sensor is put in wrist, hip left ankle and right ankle. \label{fig:errorbar}}
\end{figure}

To have a deeper look, we show the mean and standard deviation of the estimated cadence for all subjects in Figure \ref{fig:errorbar}, where we sort the subjects according to the mean cadence estimated from the hip sensor. As reported above, there is little cadence estimation discrepancy among different sensor locations when a subject is walking on the flat surface. The figure suggests that it is not guaranteed that a faster cadence of walking on the flat surface implies a faster cadence of ascending or descending stairs. However, the trend exists with statistical significance by fitting a linear regression model (that is, cadence$(i)=\beta_0+\beta_1i$, where cadence$(i)$ is the mean cadence over the walking interval and $i$ is the sorted subject index). However, the trend is less obvious when descending stairs and the trend of the estimated cadence recorded from wrist and hip is not statistically significant. The above findings are in line with our expectations.

\begin{figure}[htb!]\centering
\includegraphics[width=1\textwidth]{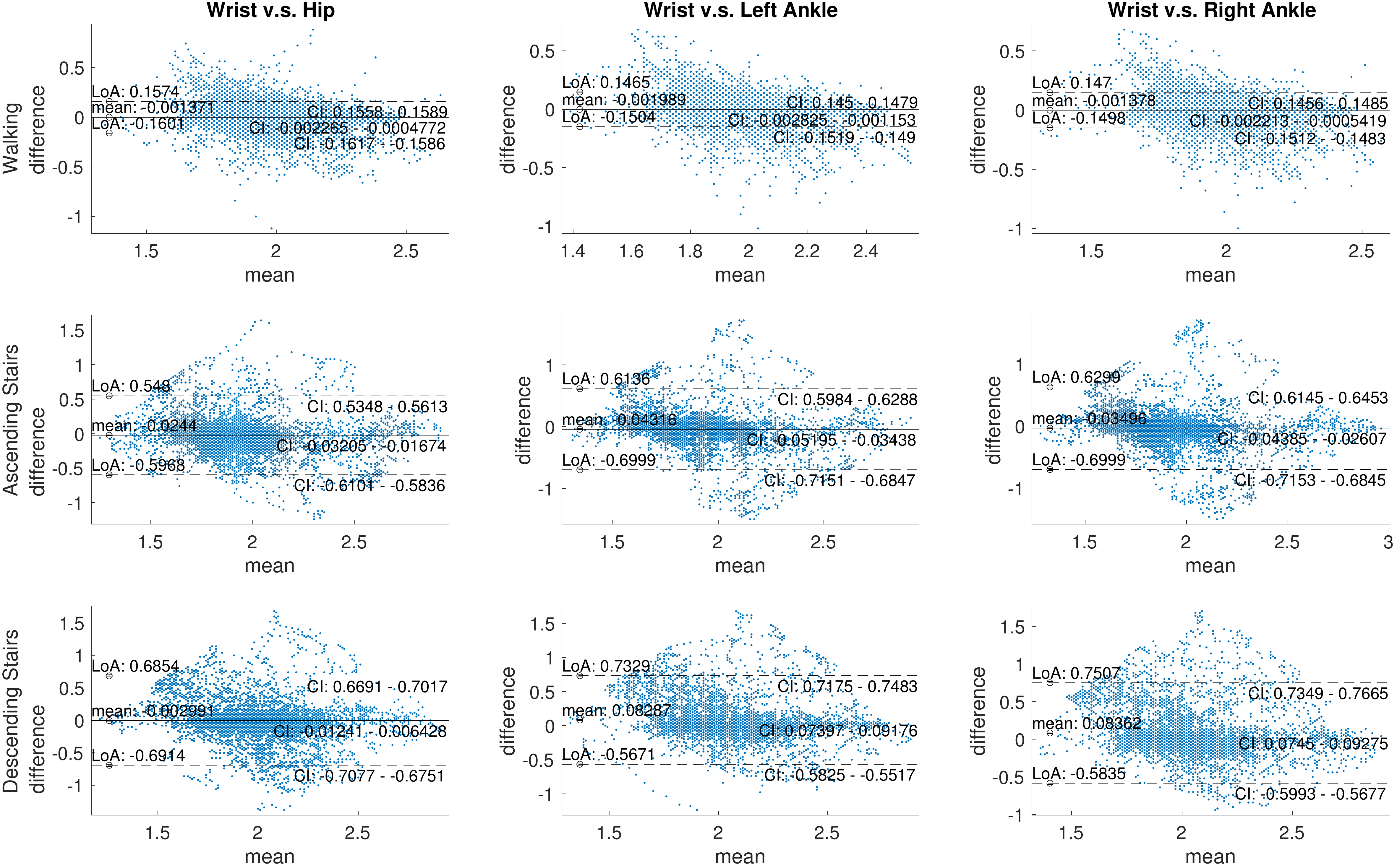}
\caption{The Bland-Altman plots when comparing different sensor locations. For left to right columns: the comparison of wrist and hip, left ankle and right ankle, respectively. From top to bottom rows: when a subject is walking, ascending stairs and descending stairs, respectively. The mean and limits of agreement (LoA) and their $95\%$ confidence intervals are shown. \label{fig:BAplot}}
\end{figure}

\begin{figure}[htb!]\centering
\includegraphics[width=1\textwidth]{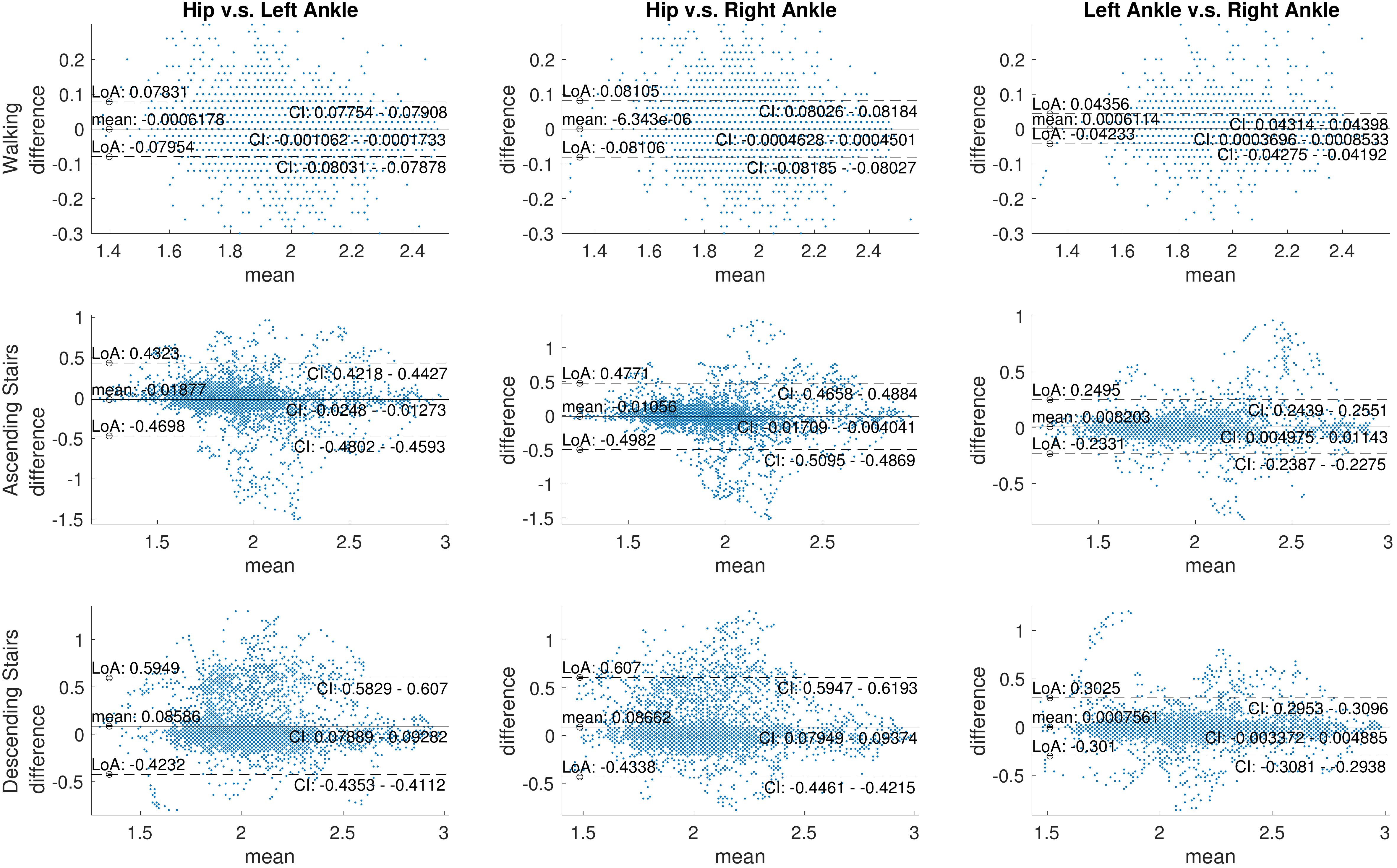}
\caption{The Bland-Altman plots when comparing different sensor locations. The caption is the same as that for Figure \ref{fig:BAplot}, except that for left to right columns: the comparisons of hip and left ankle, hip and right ankle, and left ankle and right ankle respectively. \label{fig:BAplot2}}
\end{figure}

To further study if cadence can be reliably estimated from a single sensor placed in different locations, we apply the Bland-Altman plot analysis. See Figures \ref{fig:BAplot} and \ref{fig:BAplot2} for a comparison. We see that no matter where we put the sensor, overall the agreement rate is accurate up to two digits in the mean, and the limits of agreement (LoA) are between $-0.6$ and $0.6$. The comparison between the wrist and other locations shows a larger mean and wider LoA for non-wrist locations. When comparing the results from the two ankles, the mean and LoA are both smaller. We also see that when a subject is ascending or descending stairs, the LoA is larger compared with when the subject is walking. We suggest that we could accurately and reliably estimate the cadence up to the sub-second level, while the agreement is worse when a subject is climbing stairs.

\begin{table}[hbt!]
\centering
\begin{tabular}{|c|c|c|c|}
\hline
\diaghead(3,-1){aaaaaaaaaaaaa}{Location}{Motion} & walking & ascending stairs${}^{**}$ & descending stairs${}^{*}$  \\
 \hline\hline
Wrist & 1.982$\pm$0.158 & 1.959$\pm$0.328 & 2.006$\pm$0.333   \\
\hline
Hip & 1.981$\pm$0.146 & 1.935$\pm$0.286 & 2.003$\pm$0.279 \\
\hline
Left Ankle & 1.980$\pm$0.142 & 1.916$\pm$0.263 & 2.089$\pm$0.274 \\
\hline
Right Ankle & 1.981$\pm$0.142 & 1.924$\pm$0.276 & 2.09$\pm$0.273 \\
\hline
\end{tabular}
\caption{Summary statistics of cadence estimation during different motions. The unit is steps-per-second. All values are shown as mean $\pm$ standard deviation. ${}^*$: the p-value of running the one-way ANOVA is less than $10^{-7}$; ${}^{**}$: the p-value of running the one-way ANOVA is less than $10^{-10}$.}
\label{table0}
\end{table}

\section{Discussion}\label{section:discussion}

In this work, we propose to apply dsSST to accelerometer data collected during walking to reduce non-sinusoidal oscillation in the TFR and estimate the fundamental oscillatory component. We then linked this component to a stride-to-stride frequency and obtained estimates of instantaneous gait cadence. Utilized technique is a de-shape extension of STFT aiming to handle noisy and non-stationary characters of bio-signals, which is specifically designed to alleviate the impact of harmonics and enhance the fundamental component of the biological signal that does not oscillate sinusoidally \cite{su2017extract,lu2019recycling}.

Application of the proposed cadence estimation method to the IUWDS database shows that the estimated cadence during the predefined walking intervals agrees with the intuition that descending stairs is faster than walking on the flat surface, which in turn is faster than ascending stairs. The proposed technique handles the inevitable time-varying magnitude and cadence and performs well even when walking is interrupted by rest breaks and when the accelerometer signal is recorded from the wrist.

In this work, we suggest to apply the comb filter based approach proposed by one of the authors \cite{urbanek_prediction_2018} to detect walking bout. From the statistical perspective, walking detection is equivalent to detecting $I_i$. This could be viewed as a special case of the change point detection problem. In the statistical literature, usually it is the change point of the trend that is detected. However, what we encounter in the walking detection problem is detecting the change point point of an oscillatory component while facing the time-varying nature of cadence, as well as the variation of wave-shape functions associated with the walking patterns. To our knowledge, the only work in detecting the change point of an oscillatory component is \cite{zhou2020frequency}, and it is the sinusoidal oscillatory component with fixed amplitude and frequency that is considered. How to generalize such change point detection algorithm to the walking detection problem is under study and will reported in our future work.

In short, we believe that the greatest strength of the presented work is its focus on a single sensor measurements and good performance even for wrist-worn devices. It is potential to be implemented in the real-life setting to enhance subjects' compliance. When combined with the easy-to-access, inexpensive, and multipurpose properties of the accelerometer, like the wrist-based accelerometry measurements, the proposed algorithm might open a new path toward future digital health environment.

Besides the above encouraging results, we also recognize several limitations of our approach. First, data were collected on a relatively small sample of healthy adults, which warrants further validation efforts. Also, while the experiment was designed to best mimic outdoor and free-living walking environment, using real-world and multi-day data is necessary to fully understand the utility of our algorithm. 
From the statistical perspective, the noise is in general not possible to be stationary. A theoretical justification of the algorithm under the non-stationary noise warrants a further exploration. Also, while we propose to choose $K$ in \eqref{eq:SST_recon_formula2} by a grid search approach, how to develop a more principled and faster approach would be an interesting topic that we will explore and report in our future work.

\section{Acknowledgments}

The authors thank Dr. Jacek Urbanek for fruitful discussion and the initial participation of this manuscript preparation.
The research of Jaroslaw Harezlak was partially supported by the NIMH grant R01MH108467. 

\section{Author Contributions}
HTW and JH: idea, literature review, data analysis and write-up.

\section{Competing Interests statement}
No competing interests.

\section{Reference}
\bibliographystyle{amsplain}
\bibliography{Wristcadence.bib}    

\section{Appendix}
\subsection{Technical details}

In this subsection, we provide some technical reasons for the proposed algorithm. To avoid distracting from the main idea, we only describe the basic ideas of the de-shape methodology and refer the interested reader to \cite{lin2016waveshape} for rigorous mathematical details of dsSTFT, and \cite{DaLuWu2011,Chen_Cheng_Wu:2014} for the squeezing part of dsSST. 
STFT is a widely applied tool aiming to analyze non-stationary signals. Basically, STFT divides the observed signal into pieces, analyzes the spectrum of each piece separately, and patches those spectral results together. Specifically, for a given signal $f$, with a chosen window function $h$, such as a Gaussian function centered at the origin, the STFT is defined as
\begin{equation}
V^{(h)}_f(t, \xi) = \int f(\tau) h(\tau-t)e^{-i2\pi \xi (\tau-t)} \ud \tau\,,\label{eq: stft1}
\end{equation}
where $t\in\mathbb{R}$ indicates time and $\xi\in\mathbb{R}$ indicates frequency. We call $|V^{(h)}_f(t, \cdot)|^2$ the spectrogram of the signal $f$ at time $t$, since it represents the power spectrum of the truncated signal $f(\cdot) h(\cdot-t)$ around $t$. 
Note that when the wave-shape function is non-sinusoidal, as was explained in Section \ref{Section: model} and shown in Figures \ref{fig2} and \ref{fig3}, there is an oscillatory pattern in the power spectrum $|V^{(h)}_f(t, \cdot)|^2$ caused by the harmonics. 
A naive idea leading to the cepstrum idea \cite{oppenheim2004frequency} is that the frequency of this oscillation provides information about the period of the signal $f$. To handle the time-varying frequency and amplitude nature of the biomedical signals, the {\em short-time cepstral transform (STCT)} was proposed in \cite{lin2016waveshape} and is  defined by
\begin{equation}
C^{(h,\gamma)}_f(t, q) := \int |V^{(h)}_f(t, \xi)|^\gamma e^{-i2\pi q \xi} \ud \xi,
\label{eq: rceps1}
\end{equation}
where $\gamma>0$ is sufficiently small and $q\in\mathbb{R}$ is called the quefrency (its unit is seconds or any feasible unit in the time domain). The reason for taking the $\gamma$\textsuperscript{th} power of $|V^{(h)}_f(t, \xi)|$ is simple. 
While $|V^{(h)}_f(t, \xi)|^2$ does oscillate, the amplitude of this oscillation changes from one cycle to another. To remove the influence of this amplitude modulation, we can take the natural logarithm of $|V^{(h)}_f(t, \xi)|$ so that the amplitude modulation is decoupled as a ``low-frequency component.'' However, taking the natural logarithm might be numerically unstable , so we use the approximation $ |V^{(h)}_f(t, \xi)|^\gamma$, called the ``soft logarithm.'' Systematic exploration of this idea can be found in \cite{lin2016waveshape}.
Ultimately, we obtain the fundamental {\em period} and its multiples in $C^{(h,\gamma)}_f(t, \cdot)$.

The main step in the de-shape algorithm is the inverse STCT (iSTCT), which takes into account the inverse relationship between two main quantities describing oscillation, the period and frequency:
\begin{equation}
U_f^{(h,\gamma)}(t,\xi):=C_f^{(h,\gamma)}(t,1/\xi),\label{Definition:NonlinearMask}
\end{equation}
where $\xi>0$ is given in $\mathrm{Hz}$. 
Since $C^{(h,\gamma)}_f(t, \cdot)$ captures the fundamental period and its multiples at time $t$, $U_f^{(h,\gamma)}(t,\cdot)$ captures the fundamental frequency and {\em its divisions}.
Since the common information in $V^{(h)}_f(t,\cdot)$ and $U^{(h,\gamma)}_f(t, \cdot)$ is now the fundamental frequency, we can view $U^{(h,\gamma)}_f(t, \cdot)$ as a nonlinear ``mask'' and remove multiples from the STFT by
\begin{equation}
\label{eq:W}
W^{(h,\gamma)}_f(t, \xi) := V^{(h)}_f(t,\xi)U^{(h,\gamma)}_f(t, \xi),
\end{equation}
where $\xi>0$ is interpreted as frequency. The final TF representation is $|W^{(h,\gamma)}_f(t, \xi)|^2$, which is a nonlinearly filtered spectrogram. This step can be viewed as applying a nonlinear filter to the signal to remove the influence of the wave-shape function. See Figure \ref{fig2} for the squared magnitude of $W^{(h,\gamma)}_f$. It is clear that the multiples associated with the non-sinusoidal oscillation disappear, and the TFR suggests the cadence.

If sharpening the time-frequency representation (TFR) could help, we can further consider the optional step by taking the synchrosqueezing transform (SST) idea \cite{DaLuWu2011} into account.
This nonlinear operator is produced by applying the {\em reassignment rule} \cite{DaLuWu2011,Chen_Cheng_Wu:2014}. Consider
\begin{equation}\label{definition:SSTW}
SV^{(h,\gamma,\upsilon)}_{f}(t,\xi)=\int V^{(h,\gamma)}_f(t,\eta) \delta_{|\xi-\Omega^{(h,\upsilon)}_f(t,\eta)|}\ud \eta\,,
\end{equation}
where $\xi\geq0$ and $\delta$ is the Dirac measure, and the \textit{reassignment rule} $\Omega^{(h,\upsilon)}_f$ is determined by
\begin{equation}
\Omega^{(h,\upsilon)}_f(t,\xi):=
\left\{
\begin{array}{ll}
-\Im\frac{V_f^{(\mathcal{D}h)}(t,\xi)}{2\pi V_f^{(h)}(t,\xi)}&\mbox{ when }|V_f^{(h)}(t,\xi)|> \upsilon\\
-\infty&\mbox{ when }|V_f^{(h)}(t,\xi)|\leq \upsilon.
\end{array}
\right.
\,\label{RM:omega}
\end{equation}
Here, $\mathcal{D}h(t)$ is the derivative of the chosen window function $h$, $\Im$ denotes the imaginary part, and $\upsilon>0$ gives a threshold so as to avoid instability in the computation when $|V^{(h)}_f(t,\xi)|$ is small. $SV^{(h,\gamma,\upsilon)}_{f}$ is called the SST, and $SV^{(h,\gamma,\upsilon)}_{f}(t,\cdot)$ provides a sharpened spectrogram of the oscillatory signal at time $t$. We refer interested readers to a non-mathematical tutorial \cite{wu2021new}, and several recent developments \cite{Chen_Cheng_Wu:2014,sourisseau2019inference}. 
The deSTFT can be sharpened by
\begin{equation}\label{definition:SSTW}
SW^{(h,\gamma,\upsilon)}_{f}(t,\xi)=\int |W^{(h,\gamma)}_f(t,\eta)|^2 \delta_{|\xi-\Omega^{(h,\upsilon)}_f(t,\eta)|}\ud \eta\,,
\end{equation}
where $\xi\geq0$ and $\delta$ denotes the Dirac measure, and $\Omega^{(h,\upsilon)}_f$ is the reassignment rule determined in \eqref{RM:omega}.

\end{document}